\documentclass{iau}
\usepackage{graphicx}

\title[CMEs and solar wind mass flux] 
{Solar cycle variation of coronal mass ejections contribution to solar wind mass flux}

\author[Mishra et al.]   
{Wageesh Mishra$^1$, Nandita Srivastava$^2$, Zavkiddin Mirtoshev$^3$ \and Yuming Wang$^1$ }
\affiliation{$^1$ CAS Key Laboratory of Geospace Environment, University of Science and Technology of China, Hefei, Anhui-230026, China, email: {\tt wageesh@ustc.edu.cn}
\\[\affilskip]
$^2$Udaipur Solar Observatory, Physical Research Laboratory, Udaipur-313001, India
\\[\affilskip]
$^3$Department of Physics, Samarkand State University, Samarkand- 140104, Uzbekistan}

\pubyear{2018}
\volume{340}  
\jname{Long-Term Datasets for the Understanding of Solar and Stellar Magnetic Cycle}
\editors{Dipankar Banerjee, Jie Jiang, Kanya Kusano, \& Sami Solanki}

\begin{document}
\maketitle
\begin{abstract}
Coronal Mass Ejections (CMEs) contributes to the perturbation of solar wind in the heliosphere. Thus, depending on the different phases of the solar cycle and the rate of CME occurrence, contribution of CMEs to solar wind parameters near the Earth changes. In the present study, we examine the long term occurrence rate of CMEs, their speeds, angular widths and masses. We attempt to find correlation between near sun parameters, determined using white light images from coronagraphs, with solar wind measurements near the Earth from in-situ instruments. Importantly, we attempt to find what fraction of the averaged solar wind mass near the Earth is provided by the CMEs during different phases of the solar cycles. 
\end{abstract}

\firstsection 
\section{Introduction}
Coronal Mass Ejections (CMEs) are important for our understanding of solar corona and heliosphere as they carry a huge amount of magnetized plasma from the Sun to the near Earth environment. They largely drive the space weather activity which varies almost in phase with solar cycles. Earlier studies have attempted to quantify the contribution of CMEs to background solar wind mass flux wherein the contribution of CMEs ranged from 3 to 16\% (\cite[Hildner et al. 1977]{Hildner1977}, \cite[Howard et al. 1985]{Howard1985}, \cite[Jackson \& Howard 1993]{Jackson1993}, \cite[Lamy et al. 2017]{Lamy2017}). For correct utilization of data of CME occurrence rate, the data should be corrected under consideration of instrument-dependent effects, mass and geometrical distributions of CMEs. Earlier studies were based on the shorter period of data from different coronagraphs (e.g., Skylab, Solwind, and SMM) and thus involved different duty cycle corrections and inter-calibration of visibility function. The CMEs observations has greatly improved over last 2 decades with the advent of SOHO/LASCO, and we can study their long term patterns. In our study, we reexamine the CMEs contribution using their homogeneous datasets obtained from a single instrument, the Large Angle and Spectroscopic Coronagraph (LASCO) onboard SOlar and Heliospheric Observatory (SOHO) spacecraft \cite[Brueckner et al.(1995)]{Brueckner1995}, for relatively longer period of time during
 solar cycle 23 and 24.     

\section{Methodology and Analysis}
The occurrence rate of CMEs, their angular size, speeds and mass are based on LASCO onboard  SOHO observations, as catalogued on CDAW (https://cdaw.gsfc.nasa.gov ). The occurrence rate of Interplanetary counterparts of CMEs (ICMEs) near the Earth and their speeds are based on the in in-situ observations of ACE and WIND, as compiled by Richardson \& Cane (http://www.srl.caltech.edu/ACE/ASC/DATA/level3/icmetable2.htm). We have also calculated the contribution of CMEs to the solar wind mass flux measured at 1 AU near the Earth and its variation over the solar cycle 23 and 24. For this purpose, we calculated the annual average of CME mass per day ejected into the equatorial latitude bin of 10$^\circ$ width by making appropriate corrections for CMEs latitude and their angular sizes, following the method of \cite[Howard et al. (1985)]{Howard1985}. Assuming that mass ejected into the ecliptic plane is distributed uniformly at all the longitudes, we determined the equatorial mass flux at 1 AU. Further, assuming that helium constitutes 10\% of this mass, we determined the proton flux at 1 AU due to CMEs. We noted that the solar wind flux at 1 AU and compared with the average near-ecliptic CMEs flux on an annual basis over the solar cycle. 

\begin{figure}[!h]
	\centering
	  \includegraphics[scale=0.75]{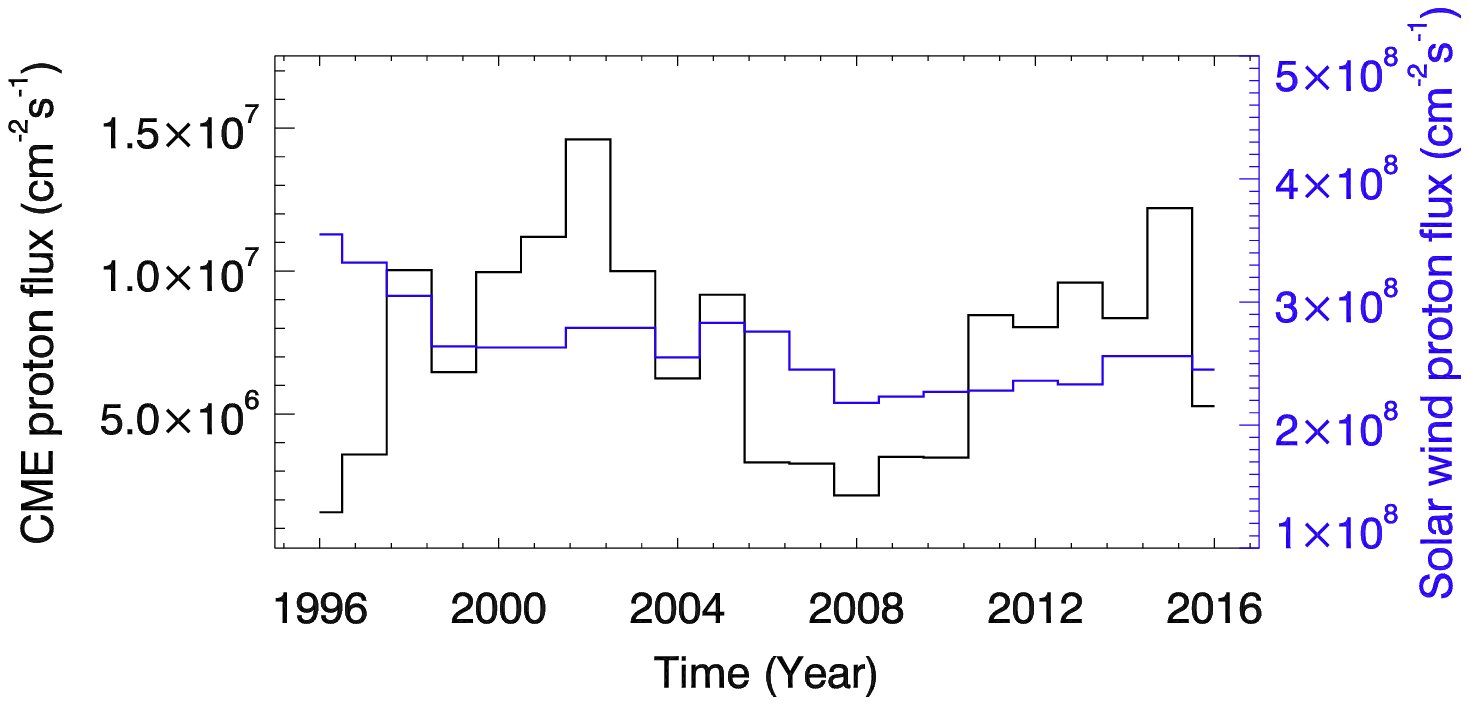}
		\hspace{0.3cm}
		\includegraphics[scale=0.75]{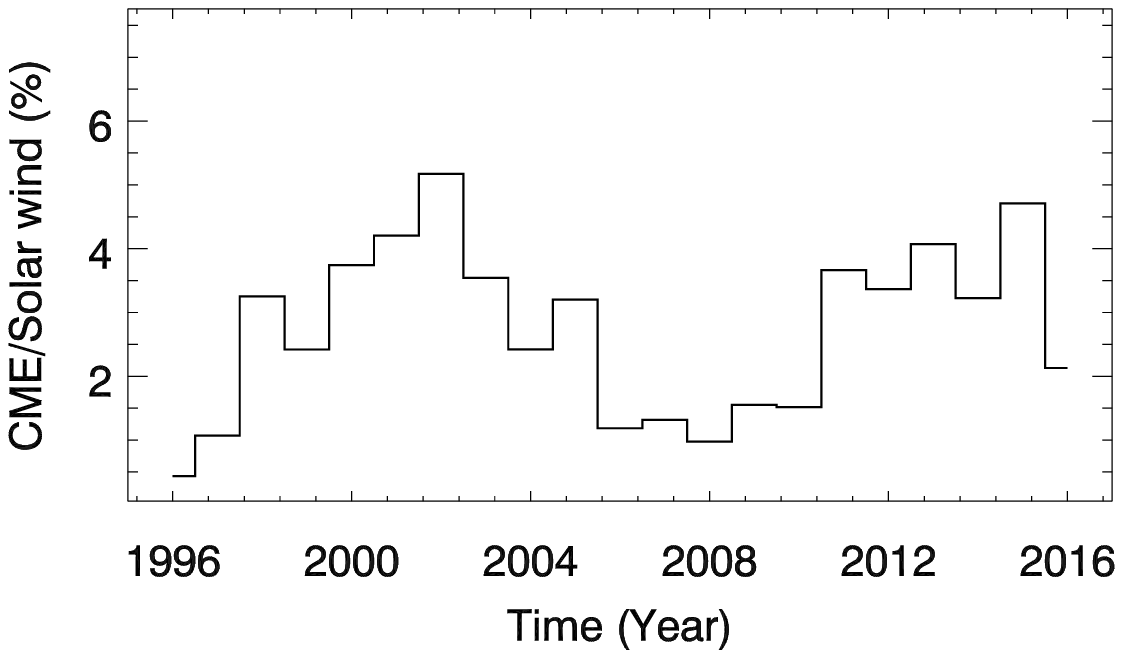}
		\caption{The variation of the CMEs and solar wind proton fluxes at 1 AU in the near-ecliptic region is shown in the top panel. The ratio of CME to solar wind mass flux is shown in the bottom panel.}
	\label{fig_CMEswflux}
\end{figure}

\section{Results and Discussion}
Based on our analysis as aforementioned, we find that in the ecliptic region, the contribution of CMEs to the solar wind mass flux is negligibly small during the solar minimum but increased to $\approx$5\% at the maximum of solar cycles 23 and 24. It is also noted that the fractional contribution of CMEs to the solar wind mass flux closely tracks the solar cycle. In ecliptic near the Earth, averaged solar wind flux is relatively constant compared to the CME flux during different phases of the solar cycle 23 and 24 (Figure~\ref{fig_CMEswflux}).

The analysis also shows that although the occurrence rate of CMEs is more for Solar cycle 24 than 23, the speeds of CMEs and ICMEs are lower in solar cycle 24 than those during solar cycle 23. The occurrence rate of CMEs tends to track the solar activity cycle 23 and 24 in both amplitude and phase except in the descending phase of cycle 23. Although sunspot numbers in solar cycle 24 are half of that in previous cycle, the rate of CMEs is little higher in solar cycle 24 than that in the corresponding phase of previous cycle 23. This may be possible because of inclusion of many faint events after the middle of solar cycle 23. In contrast to the total number, the total mass of CMEs is dominated mainly by larger events and we find that the mass loss rate at any solar latitude is lesser for solar cycle 24 than cycle 23. It is possible that fast and massive CMEs may show a stronger dependence on the sunspot numbers than weaker CMEs. The CMEs in solar cycle 24 are noted to be slower on average having a narrow speed range than those during the solar cycle 23. During the solar maximum, CMEs are wider and more uniformly distributed in latitude than during the solar minimum. 

For calculating mass flux of CMEs, we have made several approximations regarding structure of CMEs, their masses and distributions. Our approach to calculate latitude of CMEs from their central position angle (CPA) as measured from coronagraphic images may not be perfectly valid for all the CMEs. Further study is required to assess the consequence of such approximations on the estimated  CMEs contribution to solar wind mass flux.   

\acknowledgments
W.M. is supported by the Chinese Academy of Sciences (CAS) Presidents International Fellowship Initiative (PIFI) grant No. 2015PE015 and National Natural 
Science Foundation of China (NSFC) grant No. 41750110481.

\end{document}